\documentclass[12pt]{article}
\usepackage{amsmath}
\usepackage{graphicx,psfrag,epsf}
\usepackage{enumerate}
\usepackage{natbib}
\usepackage{url} 

\usepackage{natbib,bm}
\usepackage{amsmath,amssymb}
\usepackage{graphicx}
\usepackage{rotating}
\usepackage{multirow}
\usepackage{multicol}
\usepackage[small]{caption}
\usepackage{subfig}
\usepackage{color}
\usepackage{mathtools}
\usepackage{enumerate}
\usepackage{pdflscape}
\usepackage{longtable}
\usepackage{pifont}

\usepackage{setspace}

\newcommand{\nmathbf}{\bm}

\def\bfV{\nmathbf V}

\def\bfY{\nmathbf Y}

\def\bfw{\nmathbf w}
\def\bfx{\nmathbf x}

\def\bfz{\nmathbf z}

\def\bfalpha  {\nmathbf \alpha}
\def\bfbeta   {\nmathbf \beta}
\def\bfgamma  {\nmathbf \gamma}
\def\bfdelta  {\nmathbf \delta}

\def\bfmu     {\nmathbf \mu}

\def\bfphi    {\nmathbf \phi}

\newcommand{\bfzero}{{\nmathbf 0}}
\newcommand{\bfone}{{\nmathbf 1}}

\def\boldfacefake#1{\kern-4pt
   \hbox{ \mathsurround=0pt
   \hbox to 0.4pt{$#1$\hss}\hbox to 0.4pt{$#1$\hss}\hbox {$#1$}}}

\newcommand{\btable}{\begin{table}[h]\centering}
\newcommand{\etable}{\end{table}}
\newcommand{\bt}{\begin{parag}\small \let\b=\nsb \let\sb=\nssb \begin{tabular}}
\newcommand{\et}{\end{tabular}\let\b=\nb \let\sb=\nsb\end{parag}}

\newenvironment{parag}{\par}{\par}

\newcommand{\be}{\begin{eqnarray}}
\newcommand{\ee}{\end{eqnarray}}
\newcommand{\ba}{\begin{eqnarray*}}
\newcommand{\ea}{\end{eqnarray*}}

\newcommand{\reals}{\mbox{\rm I\kern-.20em R}}
\newcommand{\sreals}{\mbox{\small \rm I\kern-.20em R}}

\newcommand{\blind}{0}

\begin{document}

\def\spacingset#1{\renewcommand{\baselinestretch}%
{#1}\small\normalsize} \spacingset{1}


\if0\blind
{
  \title{\bf Bayesian Variable Selection for Multivariate Zero-Inflated Models: Application to Microbiome Count Data}
  \author{KYU HA LEE\thanks{
    To whom correspondence should be addressed. This work was supported by NIH grant CA134294, ES000002, HD052104, HD052102.}\hspace{.2cm}\\
    The Forsyth Institute, Cambridge, Massachusetts, U.S.A.\\
    and \\
    BRENT A. COULL \\
    Harvard T.H. Chan School of Public Health, Boston, Massachusetts, U.S.A.\\
    and \\
    ANNA-BARBARA MOSCICKI \\
    University of California, Los Angeles, California, U.S.A\\
    and \\
    BRUCE J. PASTER \\
    The Forsyth Institute, Cambridge, Massachusetts, U.S.A.\\
     and \\
    JACQUELINE R. STARR \\
    The Forsyth Institute, Cambridge, Massachusetts, U.S.A.}
  \maketitle
} \fi

\if1\blind
{
  \bigskip
  \bigskip
  \bigskip
  \begin{center}
    {\LARGE\bf Title}
\end{center}
  \medskip
} \fi

\bigskip
\begin{abstract}
Microorganisms play critical roles in human health and disease. They live in diverse communities in which they interact synergistically or antagonistically. Thus for estimating microbial associations with clinical covariates, such as treatment effects, joint (multivariate) statistical models are preferred.  Multivariate models allow one to estimate and exploit complex interdependencies among multiple taxa, yielding more powerful tests of exposure or treatment effects than application of taxon-specific univariate analyses. Analysis of microbial count data also requires special attention because data commonly exhibit zero inflation, i.e. more zeros than expected from a standard count distribution. To meet these needs, we developed a Bayesian variable selection model for multivariate count data with excess zeros that incorporates information on the covariance structure of the outcomes (counts for multiple taxa), while estimating associations with the mean levels of these outcomes. Though there has been much work on zero-inflated models for longitudinal data, little attention has been given to high-dimensional multivariate zero-inflated data modeled via a general correlation structure. Through simulation, we compared performance of the proposed method to that of existing univariate approaches, for both the binary (``excess zero") and count parts of the model. When outcomes were correlated the proposed variable selection method maintained type I error while boosting the ability to identify true associations in the binary component of the model. For the count part of the model, in some scenarios the univariate method had higher power than the multivariate approach. This higher power was at a cost of a highly inflated false discovery rate not observed with the proposed multivariate method. We applied the approach to oral microbiome data from the Pediatric HIV/AIDS Cohort Oral Health Study and identified five (of 44) species associated with HIV infection.
\end{abstract}


\spacingset{1.45} 

\section{Introduction}

The human microbiome plays a critical role in maintaining health and causing both acute and chronic disease. Microbes live in communities in which multiple species establish synergistic and antagonistic relationships \citep{pflughoeft2012human}. These interactions allow some species to thrive and keep others in check. The complex biological dependencies among taxa demand statistical methods that account for and exploit this interdependence. There are valid and powerful methods for jointly analyzing microbiome sequence data as predictors of health outcomes, but there are fewer methodologic options for analyzing microbiome community data as a set of joint endpoints. We specifically address three challenges that commonly arise in analysis of microbiome sequencing data as responses (dependent variables): excess zeros, interdependence of the endpoints, and the need for outcome-specific covariate selection. 

First, in most human microbiome studies, a large proportion of microbial taxa is absent in the majority of subjects, leading to many more zero counts for each taxon than expected on the basis of a Poisson, negative binomial, or Dirichlet-multinomial distribution (e.g., see Supplementary Material A) \citep{chen2013variable}. Application of a conventional linear model that uses untransformed or logarithmic-transformed counts is inappropriate for zero-inflated data \citep{loeys2012analysis, xu2015assessment}. An intuitive approach to analyzing zero-inflated count data is to view the data as arising from an underlying zero-inflated distribution, which is a mixture of a point mass at zero and a count distribution, such as Poisson \citep{lambert1992zero}. 

Second, as mentioned above, microbiome sequencing data are typically multivariate (joint response) count data sampled from communities of interdependent species. Na\"{i}ve application of a univariate, taxon-by-taxon approach implicitly assumes that counts of each taxon are uncorrelated. Although one could control for the type I error in this approach, this generally results in loss of power \citep{breiman1997predicting, la2012hypothesis}. Multivariate non-parametric methods are available compare bacterial community composition between two groups \citep{mantel1967detection, mantel1970technique, anderson2001new}; these are generally less powerful than regression methods and often do not quantify the magnitude of group differences. One approach for joint modeling of multivariate microbial sequence count data is Dirichlet-multinomial regression \citep{holmes2012dirichlet, la2012hypothesis, chen2013variable, wadsworth2017integrative}. However, the Dirichlet-multinomial model imposes restrictions that may misrepresent features of multivariate taxa count data distributions. For example, despite that relationships among microbial species can be either positively or negatively correlated, the dependence between Dirichlet variates is always negatively correlated \citep{aitchison1989multivariate, li2015microbiome}. 

Multivariate zero-inflated regression models can address both excess zeros as well as interdependent responses. Such methods that have been developed to date have been scaled to model only a small number of interdependent count endpoints, which include bivariate \citep{arab2012semiparametric, fox2013multivariate} and trivariate \citep{li1999multivariate} zero-inflated Poisson models. In some cases, existing methods have incorporated a restrictive covariance structure among outcomes, which may not always be appropriate. Specific examples of such restrictions include zero-inflated models for longitudinal data only with variance components (with no covariance components) \citep{lee2006multi, hall2000zero, leann2015marginalized}, models with dependence structures specific to spatial-temporal data  \citep{earnest2007evaluating, fernandes2009modelling, wang2015bayesian}, and models including latent factors that can induce only positive correlations among outcomes \citep{neelon2017lzip}. 

A third impediment to developing and applying a multivariate analysis technique to microbiome data is that due to having more than one endpoint, there is a large number of potential covariate-endpoint associations to be modeled. It is well recognized that variable selection helps improve prediction accuracy and reduce the cost of measurement and storage of future data. The need for variable selection techniques is well appreciated for high-dimensional covariate data and may be less well known in the context of multiple outcomes. Although there exist variable selection methods for multivariate normal \citep{brown2002multivariate, lee2017multivariate} or multinomial responses \citep{wadsworth2017integrative}, we know of no such technique applied to methods for multivariate zero-inflated outcomes.
 
We have developed multivariate zero-inflated regression models by relaxing requirements regarding the covariance structure and incorporating a Bayesian variable selection approach. The proposed methods can be used to identify zero-inflated count outcomes associated with a set of covariates while accounting for the covariance structure of the outcomes. Since it is implausible that all outcomes are relevant to the same subset of covariates, we enable the proposed model to perform outcome-specific variable selection, i.e., to identify exposures or treatments associated with particular outcomes, in this case microbial taxa. Spike-and-slab approaches have been widely used for Bayesian variable selection\citep{george1993variable, george1997approaches}, including for multivariate linear regression problems \citep{brown2002multivariate, lee2017multivariate}. In this work, we extend the spike-and-slab approach to the context of multivariate zero-inflated data.

We use the newly developed model to analyze data from the Pediatric HIV/AIDS Cohort Oral Health Study (PHACS). PHACS is an ongoing prospective cohort study at 15 US clinical sites, designed to assess the health effects of HIV infection and antiretroviral therapy (ART) on youth perinatally infected with HIV (PHIV) compared with exposed but uninfected (PHEU) youth \citep{alperen2014prevalence, tassiopoulos2016following, starr2018HIV}. The data analyzed were from a cross-sectional study focused on oral health and the oral microbiome \citep{moscicki2016burden, ryder2017prevalence}. All participants were exposed to HIV perinatally, the period when they became HIV infected if at all. Emerging from the womb, it is likely that they began acquiring their oral microbiota at birth and should have had oral microbiota similar to those of adults by 3 years of age \citep{mueller2015infant, perez2017critical}. Thus, if there is a causal association, the oral bacterial sequences we measured at 10-22 years of age more likely resulted from perinatal HIV infection rather than the reverse. This is why we treat taxa's counts as endpoints and HIV as the exposure. The goals of this analysis are (i) to identify taxa associated with HIV infection; and (ii) to estimate and test the association of HIV infection with counts of the identified taxa.

The remainder of this paper is organized as follows. Section 2 describes the proposed Bayesian framework, including model formulation and specification of prior distributions. Section 3 describes results from simulation studies conducted to compare the operating performance of the proposed variable selection approach versus an existing univariate method. Section 4 describes results from the PHACS data analysis. In section 5 we further discuss the method and results.

\section{Methods}\label{sec:method}

In this section, we describe the proposed multivariate zero-inflated model, present prior distributions for model parameters and the variable selection strategy, discuss interpretation of the regression parameters, and summarize the computational scheme and implementation (see Supplementary Material B for a summary of model parameters and C for implementation details).

\subsection{Model Formulations} \label{sec:model}

Suppose that count outcomes $Y_{ij}$ are observed for taxon $j$=1,\ldots,$q$ and subject $i$=1,\ldots,$n$. We use an approach that assumes that $Y_{ij}$ follow a multivariate zero-inflated Poisson (MZIP) distribution, which is a mixture of a Poisson distribution and point mass distribution at zero ($\mathcal{I}_0$):
\be
	Y_{ij} &\sim& \textrm{Poisson}(\lambda_{ij}), ~~~~~ \textrm{if } U_{ij}=1, \nonumber  \\
		  &\sim& \mathcal{I}_0, 			~~~~~~~~~~~~~~~~~~ \textrm{if } U_{ij}=0, \label{eq:model}
\ee 
where $U_{ij}$ is an unobservable indicator for the excess zeros for taxon $j$ in subject $i$, and $\lambda_{ij}$ is the mean of the Poisson distribution. The model implies that some zeros occur through a Poisson process whereas others represent the impossibility for a given taxon to be observed in some subjects. In practice, regression analysis based on the MZIP model proceeds by placing structure on $\lambda_{ij}$ and P[$U_{ij}=1$], specifically as a function of the covariates and random effects. Toward this, let $\bfx_{i}$ and $\bfz_{i}$ be a vector of $p_x$ and $p_z$ covariates for the $i^{\textrm{th}}$ subject that will be considered in the model for $\lambda_{ij}$ and P[$U_{ij}=1$], respectively. With this formulation it is not necessary for the presence and the count of a taxon to depend on the same set of covariates. 

For the count (Poisson) model part, we consider the following general modeling specification:
\be
	\log(\lambda_{ij}) &=& \beta_{0j}+\bfx_i^{\top}\bfbeta_j + \log(\xi_i) +  V_{ij}, \label{eq:model:count}
\ee
where $\bfbeta_0$=($\beta_{01}$,\ldots,$\beta_{0q}$)$^{\top}$ are the outcome-specific intercepts and $\bfbeta_j$=($\beta_{j1}$,\ldots,$\beta_{jp_x}$)$^{\top}$ are the outcome-specific vectors of fixed-effect regression parameters. The random effects $\bfV_i=(V_{i1},\ldots,V_{iq})^{\top}$ characterize the unobserved characteristics that are associated with the mean count for taxon $j$ in subject $i$ and account for within-subject correlations. The term $\log(\xi_i)$ is included as an offset variable for settings in which one is interested in the incidence density $\lambda_{ij}/\xi_i$. For application to genetic sequence counts, setting $\xi_i$ to the total number of sequencing reads accounts for individual variation in sequencing depth.

To account for the dependency structure in the binary part of the model, we adopt a multivariate probit model \citep{ashford1970multi}. Letting $U_{ij}$=$I(w_{ij}\geq0)$, with indicator function $I(\cdot)$ , we consider a multivariate normal (MVN) distribution for the latent variable $\bfw_i$=($w_{i1}$,\ldots,$w_{iq}$)$^{\top}$, with location vector $\bfalpha_0+A^{\top}\bfz_i$ and variance-covariance matrix $R$. Here, $\bfalpha_0$=($\alpha_{01}$,\ldots,$\alpha_{0q}$)$^{\top}$ are the intercepts and $A$ is the $p_z\times q$ coefficient matrix whose columns are $\bfalpha_j$=($\alpha_{j1}$, \ldots,$\alpha_{jp_z}$)$^{\top}$, $j$=1,$\cdots$,$q$. Then the probability density function of $U_{ij}$ is given by
\be
	\bfw_i | \bfalpha_0, A, R,  \bfz_i &\sim& \textrm{MVN}_q(\bfalpha_0+A^{\top}\bfz_i, ~ R). \label{mvprobit}
\ee
$R$ is restricted to be a correlation matrix to ensure identifiability of all model parameters \citep{chib1998analysis, liu2001discussion}. $R$ measures the dependence between $U_{ij}$ and $U_{ij'}$ by using the correlations among the elements of the vector $\bfw_i$. Let $\vec{\bfY}$, $\vec{\bfw}$, $\vec{\bfV}$, and $\bfphi$ denote the collections of $\bfY_i$, $\bfw_i$, $\bfV_i$, and $\phi_i$, respectively, across all subjects. Let $B$ be the $p_x\times q$ coefficient matrix whose columns$\bfbeta_j$=($\beta_{j1}$, \ldots,$\beta_{jp_x}$)$^{\top}$, $j$=1,$\cdots$,$q$. Combining (\ref{eq:model}), (\ref{eq:model:count}), and (\ref{mvprobit}), the augmented data likelihood function, as a function of the unknown parameters $\Theta$=\{$\bfalpha_0$, $A$, $\bfbeta_0$, $B$, $\vec{\bfV}$, $R$\}, is:
\be
	L(\vec{\bfY}, \vec{\bfw}|\Theta)	&=& \prod_{i=1}^n\prod_{j=1}^q \left[ I(w_{ij}\geq0)\frac{\exp\left\{ y_{ij}\log(\lambda_{ij})-\lambda_{ij} \right\} }{y_{ij}!} +I(w_{ij}<0)I(y_{ij}=0) \right] \nonumber \\
	&\times& \prod_{i=1}^n (2\pi)^{-q/2} \left|R \right|^{-\frac{1}{2}}\exp\left\{-\frac{1}{2}(\bfw_i-\bfalpha_0-A^{\top}\bfz_i)^{\top}R^{-1}(\bfw_i-\bfalpha_0-A^{\top}\bfz_i)\right\}. \label{lh}
\ee

\subsection{Prior Specification and Covariance structure} {\label{sec:prior}}

We complete the Bayesian formulation of the proposed framework by specifying prior distributions for the unknown parameters. To facilitate outcome-specific variable selection, we adopt spike-and-slab priors for the regression parameters in both parts of the proposed model. Such a prior has been widely used in the context of Bayesian stochastic search variable selection \citep{george1993variable, george1997approaches}. Specifically, we assign the following priors for $\bfbeta_j$ and $\bfalpha_j$: 
\vspace{-0.05in}
\be
	\beta_{j,k} | \gamma_{j,k}, \sigma^2_{\beta, k} &\sim& \gamma_{j,k}\textrm{Normal}(0, v_{\beta,j}^2\sigma^2_{\beta, k}) + (1-\gamma_{j,k})\mathcal{I}_0 \textrm{ and} \nonumber \\
	\alpha_{j,l} | \delta_{j,l}, \sigma^2_{\alpha, l} &\sim& \delta_{j,l}\textrm{Normal}(0, v_{\alpha,j}^2\sigma^2_{\alpha, l}) + (1-\delta_{j,l})\mathcal{I}_0, \label{eq:spike}
\ee 
where $\bfgamma_{k}=(\gamma_{1,k}, \ldots, \gamma_{q,k})^{\top}$ and $\bfdelta_{l}=(\delta_{1,l}, \ldots, \delta_{q,l})^{\top}$, for $k=1,\ldots,p_x$, $l=1,\ldots,p_z$, are vectors 

\noindent of binary latent variables indicating the membership of each regression parameter to one of the mixture components in (\ref{eq:spike}). The $k$th covariate is considered to be associated with mean counts for the $j$-th outcome if the data support $\gamma_{j,k}=1$ over $\gamma_{j,k}=0$. A similar interpretation holds for $\delta_{j,k}$. We use independent Bernoulli hyperpriors for $\gamma_{j,k}$ and $\delta_{j,l}$ with inclusion probability $\omega_{\beta, k}$ and $\omega_{\alpha, l}$, respectively.

As outlined in Section \ref{sec:model}, the dependence among mean bacterial counts of multiple taxa is accounted for by the distribution of random effects $\bfV_i$. Specifically, we structure this dependence through a hierarchical Poisson-logNormal model \citep{aitchison1989multivariate, fox2013multivariate}, which corresponds to a MVN prior, $\bfV_i | \Sigma_V \sim \textrm{MVN}(\bfzero, \Sigma_V)$. We use an unstructured covariance pattern for $\Sigma_V$ and $R$, thus imposing no specific structure for the dependence among outcomes. Under the unstructured model, we adopt a conjugate hyperprior, inverse-Wishart($\Psi_0$, $\rho_0$), for $\Sigma_V$. For $R$, we use $\pi(R)\propto |R|^{-(q+1)/2}$, which is the prior for a correlation matrix based on Jeffreys' prior distribution for the variance-covariance matrix \citep{box2011bayesian}. In addition, we assign $\bfalpha_0\sim \textrm{MVN}(\bfmu_{\alpha_0}, \sigma^2_{\alpha_0} R)$, where $\bfmu_{\alpha_0}$ is the hyperparameter to be specified. We discuss the desirable properties of the prior distributions for $R$ and $\bfalpha_0$ in Supplementary Material C \citep{chib1998analysis, liu2000analysis}. We assign MVN($\bfmu_{\beta_0}$, $\sigma^2_{\beta_0}$$I_q$) for the intercepts $\bfbeta_0$, where $\bfmu_{\beta_0}$ is the hyperparameter to be specified, and $I_q$ is the $q$$\times$$q$ identity matrix. Lastly, we use the conjugate hyperpriors inverse-Gamma($a_{\beta, k}$, $b_{\beta, k}$), inverse-Gamma($a_{\alpha, l}$, $b_{\alpha, l}$), inverse-Gamma($a_{\beta_0}$, $b_{\beta_0}$) and inverse-Gamma($a_{\alpha_0}$, $b_{\alpha_0}$), for $\sigma^2_{\beta, k}$, $\sigma^2_{\alpha, l}$, $\sigma^2_{\beta_0}$, and $\sigma^2_{\alpha_0}$, respectively.

\subsection{Induced Marginal Incident Density Ratio} \label{sec:IDR}

Because the MZIP model is a mixture model, the regression parameters, $\beta_{j,k}$, have latent interpretations: $\beta_{j,k}$ represents the change in log mean count of taxon $j$ associated with a one-unit increase in the covariate $k$, in a susceptible sub-population \citep{preisser2012review}. The relationship between $E[Y_{ij}| V_{ij}]$ and the parameters from the proposed MZIP model is given by
\be
	E[Y_{ij}| V_{ij}] = \lambda_{ij}P(U_{ij}=1) = \exp\left(\beta_{0j}+\bfx_i^{\top}\bfbeta_j + \log(\xi_i) +  V_{ij}\right) \Phi\left(\alpha_{0j}+\bfz_i^{\top}\bfalpha_j\right), \nonumber
\ee
where $\Phi(\cdot)$ is the cumulative distribution function of the standard normal distribution. For models with log-links and normally distributed random effects such as the one we propose, it is straightforward to marginalize the conditional expectation over the random effects distribution \citep{leann2015marginalized}. Under a model with $\bfx_i=\bfz_i$, the ratio of means for a one-unit increase in covariate $k$ is given by
\be
	\textrm{IDR}(\bfx_{i,(-k)})&=&\frac{E[Y_{ij}| x_{ik}=x+1, \bfx_{i,(-k)}]}{E[Y_{ij}| x_{ik}=x, \bfx_{i,(-k)}]}= \frac{E\left[E[Y_{ij}| x_{ik}=x+1, \bfx_{i,(-k)}, V_{ij}]\right]}{E\left[E[Y_{ij}| x_{ik}=x, \bfx_{i,(-k)}, V_{ij}]\right]}\nonumber \\	
	&=& e^{\beta_{jk}}\frac{\Phi\left(\alpha_{0j} + \bfx_{i,(-k)}^{\top}\bfalpha_{j,(-k)} + (x+1)\alpha_{j,k}\right)}{\Phi\left(\alpha_{0j} + \bfx_{i,(-k)}^{\top}\bfalpha_{j,(-k)} + x\alpha_{j,k}\right)}, \label{marginalEffect}
\ee  
where $\bfx_{i,(-k)}$ and $\bfalpha_{j,(-k)}$ are the vectors $\bfx_{i}$ and $\bfalpha_{j}$, respectively, with the $k$-th element removed. Although we obtain $\textrm{IDR}(\bfx_{i,(-k)})$ by marginalizing $E[Y_{ij}| V_{ij}, U_{ij}=1]$ over the latent mixture distribution and the distribution of random effect $V_{ij}$, the quantity still depends on the values of other covariates $\bfx_{i,(-k)}$, which can be addressed several ways \citep{preisser2012review}. For continuous covariates, one could obtain a covariate-adjusted $\textrm{IDR}(\bar{\bfx}_{(-k)})$ by either i) inserting mean values of covariates $\bfx_{i,(-k)}$, or ii) assuming specific covariate distributions and marginalizing the quantity over these distributions. For discrete covariates, one could iii) empirically marginalize the $\textrm{IDR}(\bfx_{i,(-k)})$ over the observed distribution of covariates, or iv) present multiple different values for the $\textrm{IDR}(\bfx_{i,(-k)})$, one for each category defined by unique covariate profiles.

As has been well described, estimating the variance of measures such as the $\textrm{IDR}(\bfx_{i,(-k)})$ in a frequentist framework would require an extra statistical technique such as bootstrap resampling \citep{albert2014estimating}. An advantage of the Bayesian paradigm is that estimation of uncertainty for $\textrm{IDR}(\bfx_{i,(-k)})$ follows directly from the variance of its posterior distribution, estimated by evaluating its expression at each scan of the Markov chain Monte Carlo (MCMC) scheme.

\vspace{-0.2in}
\subsection{Markov Chain Monte Carlo}

We perform estimation and inference for the proposed model by using a Gibbs sampling algorithm to generate samples from the posterior distribution. In the corresponding MCMC scheme, parameters are updated either by exploiting conjugacies inherent in the model structure or by using a Metropolis-Hastings step. However, MCMC is far from straightforward because the joint posterior distribution under the proposed framework involves i) the unobserved multivariate latent variables, $\bfw_i$; ii) the augmented data likelihood function based upon the latent mixture distribution; iii) spike-and-slab mixture priors for the regression parameters; iv) a high-dimensional parameter space due to the unstructured pattern for $\Sigma_V$ and $R$; and v) restrictions on the correlation parameters in $R$. Therefore, we develop an efficient MCMC sampling scheme based on a data augmentation algorithm \citep{tanner1987calculation} in which the computational challenge of high-dimensionality is avoided by iterating between an ``imputation step," in which values of the unobserved latent variables $\bfw_i$ are imputed and updated, and a ``posterior step," in which the model parameters are updated. In the posterior step, we used a parameter expanded data augmentation method \citep{liu2001discussion} to update $R$ for computational efficiency (see Supplementary Material C for detailed description of the proposed computation scheme). We have developed a series of core functions in $\texttt{C}$ to improve the computation speed, for which we provide the algorithm in the \texttt{mBvs} package for \texttt{R} (https://cran.r-project.org/web/packages/mBvs). As an example of computational time, it takes approximately 3 minutes to generate 10,000 MCMC scans on a 2.5 GHz Intel Core i7 MacBook Pro for the analysis of the PHACS data ($n$=254, $q$=44, $p_x$=$p_z$=2).

\section{Simulation Studies} \label{sec:simulation}

We evaluated the performance of the proposed method on simulated data. We generated data sets under six scenarios with varying outcome correlation structures and association patterns between a covariate and the vector of outcomes in the two model parts.

\subsection{Set-Up and Data Generation}\label{sec:sim:setup}

We generated samples of size $n=300$ with $q=20$ outcomes and $p_x=p_z=1$ covariate under the proposed model given in (\ref{eq:model}), (\ref{eq:model:count}), and (\ref{mvprobit}). The covariate was generated from Normal(0, 2) and the intercepts set to $\bfalpha_0=\bfone_q$ and $\bfbeta_0=5\cdot\bfone_q$. In Scenarios I-III and VI, we varied the scale and sign of the association between the covariate and outcomes in the two model parts by setting  $B=A=$[0.05, 0.10, 0.15, 0.20, 0.25, -0.05, -0.10, -0.15, -0.20, -0.25, $\bfzero_{10}^{\top}$]. In Scenario IV, the covariate was associated with outcomes in only one of the two model parts: $B=$[0.05, 0.10, 0.15, 0.20, 0.25, $\bfzero_{15}^{\top}$] and $A=$[$\bfzero_{5}$, 0.05, 0.10, 0.15, 0.20, 0.25, $\bfzero_{10}^{\top}$]. We considered the \emph{null} case in Scenario V by setting all elements of $B$ and $A$ to zero. We set each variance-covariance matrix $\Sigma_V$ in the count part of the model and $R$ in the binary part of the model to a correlation matrix with an exchangeable structure with correlation $c_1$ within the block of the first ten outcomes, an exchangeable structure with correlation $c_2$ within the block of the second 10 outcomes, and a common cross-block correlation of $c_3$ for pairs of outcomes from different blocks. In Scenarios I, IV, and V, the outcomes associated with the covariate were highly correlated and outcomes unassociated with the covariate only moderately correlated, ($c_1$, $c_2$, $c_3$)=(0.70, 0.30, 0.20). In Scenario II, the outcomes associated with the covariate were moderately correlated and the remaining outcomes weakly correlated, ($c_1$, $c_2$, $c_3$)=(0.40, 0.05, 0.10). In Scenario III, the outcomes associated with the covariate were weakly correlated and those unassociated with the covariate highly correlated, ($c_1$, $c_2$, $c_3$)=(0.20, 0.70, 0.30). In Scenario VI, each outcome is assumed to follow a univariate zero-inflated Poisson (UZIP) distribution, indicating ($c_1$, $c_2$, $c_3$)=(0, 0, 0) (independence), and diag($\Sigma_V$)=diag(R)=$\bfzero$ (no overdispersion). 

\subsection{Analyses and Specification of Hyperparameters}\label{sec:sim:hyperparameter}

We fit the proposed MZIP model to 600 simulated data sets, 100 under each of the six scenarios. For comparison purposes in each data set, we also implemented UZIP regression with and without the lasso penalty by using the \texttt{pscl}\citep{zeileis2008regression} and   \texttt{mpath}\citep{wang2015variable} \texttt{R} packages, respectively. As outlined in Section \ref{sec:prior}, the proposed Bayesian framework requires the specification of several hyperparameters. For the intercepts $\bfbeta_0$ and $\bfalpha_0$ and their variance components $\sigma_{\alpha_0}^2$ and $\sigma_{\beta_0}^2$, we assigned noninformative priors by setting ($\bfmu_{\beta_0}$, $a_{\beta_0}$, $b_{\beta_0}$)=($\bfmu_{\alpha_0}$, $a_{\alpha_0}$, $b_{\alpha_0}$)=($\bfzero_q$, 0.7, 0.7). For the regression parameters, $\bfbeta$ and $\bfalpha$, and their variance components, $\sigma_{\alpha}^2$ and $\sigma_{\beta}^2$, the hyperparameters ($v_{\beta, j}$, $a_{\beta, k}$, $b_{\beta, k}$)=($v_{\alpha, j}$, $a_{\alpha, l}$, $b_{\alpha, l}$), $k=1,\ldots, p_x$, $l=1,\ldots, p_z$, were set to (10, 0.7, 0.7) to make the corresponding priors fairly noninformative. The hyperparameters $\omega_{\beta, k}=\omega_{\alpha, l}$ were set to either 0.1 or 0.5, implying 0.1 or 0.5 \emph{a priori} probability, respectively, of each covariate to be selected as associated with each outcome. Finally, we set $(\Psi_0, \rho_0)=(\psi I_q, q+\psi+1)$ with $\psi=3$, corresponding to a prior distribution of $\Sigma_V$ centered at $I_q$ and with variance of the diagonal elements equal to 2.0. We ran each MCMC chain for 1 million iterations with the first half taken as burn-in. 

For the proposed Bayesian MZIP model, we perform variable selection based on the marginal posterior distribution of variable selection indicators, $\gamma_{j,1}$ and $\delta_{j,1}$. Here, we applied a marginal posterior probability cutoff of 0.5. Between the two univariate approaches implemented, in initial simulation studies the penalized UZIP model tended to select the covariate for all outcomes in both model parts when the outcomes were correlated. For this reason, and because it would be one typical practice, for the UZIP regression analyses we performed variable selection by applying 95\% confidence intervals with a false discovery rate (FDR)-controlling procedure \citep{benjamini1995controlling} to account for multiple comparisons. 

We assessed performance of the variable selection feature of the model by calculating four quantities based on the true positives (TP; the number of outcomes associated with the covariate and selected into the model) and false positives (FP; the number of outcomes unrelated to the covariate and mistakenly selected into the model), where $\tau$ is the number of outcomes that are truly associated with the covariate: the true positive rate, TPR=TP/$\tau$, the false positive rate, FPR=FP/($20-\tau$), the positive predictive value, PPV=TP/(TP+FP), and the negative predictive value, NPV=($20-\tau-\textrm{FP}$)/($20-\textrm{TP}-\textrm{FP}$). 

\begin{table}[!h]
\caption{Four operating characteristics$^{*}$ (\%) and the number of outcomes selected to be associated with the covariate ($q_{\textrm{sel}}$) for the univariate zero-inflated Poisson (UZIP)$^{\dag}$ and the proposed multivariate zero-inflated Poisson (MZIP)$^{\ddag}$ models across six simulation scenarios described in Section \ref{sec:sim:setup}.} \label{sim:tbl:roc}
\centering
\scalebox{0.7}{
\begin{tabular}{c c rrrr r rrrr r rrrr}
  \hline
& &  \multicolumn{4}{c}{UZIP} &&  \multicolumn{9}{c}{MZIP} \\ \cline{3-6}  \cline{8-16}
& &  &&& && \multicolumn{4}{c}{($\omega_{\alpha}=\omega_{\beta}=0.1$)} & & \multicolumn{4}{c}{($\omega_{\alpha}=\omega_{\beta}=0.5$)} \\ \cline{8-11} \cline{13-16}
Scenario & &  \multicolumn{2}{c}{\underline{Binary}} & \multicolumn{2}{c}{\underline{Count}}  && \multicolumn{2}{c}{\underline{Binary}} & \multicolumn{2}{c}{\underline{Count}} && \multicolumn{2}{c}{\underline{Binary}} & \multicolumn{2}{c}{\underline{Count}} \\
& &  Mean & (SD) & Mean & (SD)  && Mean & (SD) & Mean & (SD) && Mean & (SD) & Mean & (SD) \\
  \hline
	&TPR 					& 54.3 & (9.7) & 99.1 & (2.9) &  	& 67.4 &(8.4) & 82.5 &(5.5) &  		& 78.4 &(8.1) & 87.4 &(5.5)  \\ 
	& FPR 					& 0.1 & (1.0) & 86.0 & (11.1)&  	& 0.3 &(1.8) & 0.8 &(2.8) &  		&  4.6 &(6.5) & 3.1 &(5.5) \\ 
I 	& PPV 					& 99.8 & (1.7) & 53.7 & (3.4)&  		& 99.6 &(2.4) & 99.1 &(3.0) &  		& 95.0 &(6.8) & 96.9 &(5.3) \\ 
 	& NPV 					& 68.9 & (4.7) & 94.4 & (17.0)&  	& 75.6 &(4.9) & 85.2 &(4.2) &  		& 81.9 &(5.6) & 88.7 &(4.4)  \\ 
 	& $q_{\textrm{sel}}$	& 5.4 & (1.0) & 18.5 & (1.2) &  		&  6.8 &(0.8) & 8.3 &(0.6)	&  		&  8.3 &(1.1) & 9.1 &(0.8)  \\ 
  \hline
  	&TPR 			& 53.7 & (9.9) & 99.3 & (2.6) &  & 60.8 & (9.6) & 75.2 & (8.0) &  	& 73.0 & (8.8) & 84.1 & (7.0) \\ 
	& FPR 			& 0.4 & (1.9) & 87.1 & (10.0) &  & 0.5 & (2.1) & 0.6 & (2.4) &  	& 4.0 & (5.8) & 2.7 & (5.1) \\ 
 II	& PPV 			& 99.5 & (2.8) & 53.4 & (2.9) &  & 99.4 & (2.8) & 99.3 & (2.8)&  	& 95.3 & (6.7) & 97.2 & (5.4)  \\ 
 	& NPV 			& 68.6 & (4.6) & 96.4 & (14.2) &  & 72.1 & (4.9) & 80.4 & (5.3) &  	& 78.4 & (5.6) & 86.2 & (5.5) \\ 
 	& $q_{\textrm{sel}}$ & 5.4 & (1.0) & 18.6 & (1.1) &  	& 6.1 & (1.0) & 7.6 & (0.8)&  		& 7.7 & (1.1) & 8.7 & (0.8)  \\ 
\hline
	&TPR 			& 56.0 & (10.2) & 99.1 & (2.8) 		&  & 60.0 & (10.0) & 73.5 & (8.3)&  & 74.3 & (9.2) & 82.6 & (7.9)  \\ 
 	& FPR 			& 0.2 & (1.5) & 88.7 & (11.1) 		&  & 0.2 & (1.5) & 0.1 & (1.0) &  & 3.9 & (5.8) & 1.8 & (4.1) \\ 
III 	& PPV 			& 99.6 & (2.5) & 53.0 & (3.3) 		&  & 99.6 & (2.4) & 99.9 & (1.2) &  & 95.2 & (7.1) & 98.1 & (4.3) \\ 
 	& NPV 			& 69.8 & (5.0) & 92.3 & (23.2) 		&  & 71.7 & (5.1) & 79.4 & (5.2) &  & 79.2 & (6.3) & 85.4 & (5.9) \\ 
 	& $q_{\textrm{sel}}$ & 5.6 & (1.0) & 18.8 & (1.2) 		&  & 6.0 & (1.0) & 7.4 & (0.8) &  & 7.8 & (0.9) & 8.4 & (1.0) \\ 
\hline
 	&TPR 				& 56.8 &(16.3) & 98.8 &(4.8) &  	& 63.0 &(17.4) & 82.6 &(11.2) &  	& 79.0 &(15.1) & 88.6 &(10.7) \\ 
 	& FPR 				& 0.1 &(0.7) & 86.8 &(8.6) &  		& 0.4 &(1.6) & 0.3 &(1.3) &  		& 3.3 &(4.6) & 2.5 &(4.2) \\ 
IV  	&PPV 				& 99.8 &(2.0) & 27.6 &(2.5) &  		& 98.5 &(6.0) & 99.3 &(3.6)&  		& 90.3 &(12.8) & 93.3 &(10.8)  \\ 
  	&NPV 				& 87.6 &(4.2) & 95.9 &(17.1) &  	& 89.2 &(4.6) & 94.6 &(3.4) &  		& 93.4 &(4.5) & 96.3 &(3.4) \\ 
  	&$q_{\textrm{sel}}$ 		& 2.9 &(0.8) & 18.0 &(1.3) &  		& 3.2 &(0.9) & 4.2 &(0.6)&  		& 4.5 &(1.0) & 4.8& (0.8)  \\ 
\hline
	&TPR 					&  - 		& - 		& - & - &  &-  &-  &-  &-  &  & - & - & -  &-  \\ 
	& FPR 					& 0.1		& (0.5) 	& 86.8	 & (8.5) 		&  & 0.1 & (0.7) & 0.1 & (0.7) &  & 1.9 & (3.4) & 1.1 & (2.3) \\ 
V$^{\S}$& PPV 					&  - 		& - & - & - &  & -  &-  &-  &-  &   & - & - & -  &-  \\ 
 	& NPV 					& 100.0 	& (0.0) 	& 100.0 & (0.0) 		&  & 100.0 & (0.0) & 100.0 & (0.0)&  & 100.0 & (0.0) & 100.0 & (0.0)  \\ 
 	& $q_{\textrm{sel}}$ 			& 0.0 	& (0.1) 	& 17.4 & (1.7) 		&  & 0.0 & (0.1) & 0.0 & (0.1) &  & 0.4 & (0.7) & 0.2 & (0.5) \\
\hline  
	&TPR 			& 58.1 &(10.7) & 100.0 &(0.0)  	&  & 57.7 & (10.8) & 100.0 & (0.0)&  	& 72.2 & (9.6) & 100.0 & (0.0)  \\ 
  	&FPR			& 0.0 &(0.0) & 0.0 &(0.0) 	&  & 0.1 & (1.2) & 0.0 & (0.0) &  	& 4.7 & (6.3) & 0.1 & (1.1) \\ 
VI  	&PPV 			& 100.0 &(0.0) & 100.0 &(0.0) 	&  & 99.8 & (1.9) & 100.0 & (0.0) &  	& 94.4 & (7.3) & 99.9 & (1.0) \\ 
  	&NPV 			& 70.9 &(5.5) & 100.0 &(0.0) 	&  & 70.6 & (5.4) & 100.0 & (0.0) &  	& 77.8 & (6.1) & 100.0 & (0.0) \\ 
  	&$q_{\textrm{sel}}$ 	& 5.8 &(1.1) & 10.0 &(0.0) 	 &  & 5.8 & (1.1) & 10.0 & (0.0) &  	& 7.7 & (1.1) & 10.0 & (0.1)\\ 
   \hline   
   \multicolumn{16}{l}{\footnotesize$^{*}$TPR=TP/$\tau$, FPR=FP/($20-\tau$), PPV=TP/(TP+FP), NPV=($20-\tau-\textrm{FP}$)/($20-\textrm{TP}$+FP), where TP is the number of outcomes} \\
   \multicolumn{16}{l}{\footnotesize ~ associated with the covariate and selected into the model, FP is the number of outcomes unrelated to the covariate and mistakenly} \\      
   \multicolumn{16}{l}{\footnotesize ~ selected into the model, and $\tau$ is the number of outcomes that are truly associated with the covariate. Note, $\tau$ is $10$ in Scenarios I-III}\\       
      \multicolumn{16}{l}{\footnotesize ~  and VI, $5$ in Scenario IV, and $0$ in Scenario V.}\\        
   \multicolumn{16}{l}{\footnotesize$^{\dag}$For variable selection in UZIP analyses, we used 95\% confidence intervals with a false discovery rate controlling procedure.}\\            
   \multicolumn{16}{l}{\footnotesize$^{\ddag}$For variable selection in MZIP models, we applied a marginal posterior probability cutoff of 0.5 for both $\gamma$ and $\delta$. The hyperparameters}\\
   \multicolumn{16}{l}{\footnotesize ~ $\omega_{\alpha}$=$\omega_{\beta}$ are set to either 0.1 or 0.5}\\
   \multicolumn{16}{l}{\footnotesize$^{\S}$Since there is no outcomes that are truly associated with the covariate in Scenario V, TPR and PPV are not presented.}\\       
   \multicolumn{16}{l}{\footnotesize NOTE: Throughout the mean and standard deviation (SD) values are based on results from 100 simulated datasets.}\\        
\end{tabular}}
\end{table}

\begin{table}[h]
\caption{Estimated regression parameters and inclusion probabilities for the univariate zero-inflated Poisson (UZIP)$^{\dag, *}$ and the proposed multivariate zero-inflated Poisson (MZIP)$^{\ddag}$ models for Scenario I.} \label{sim:tbl:est}
\centering
\scalebox{0.8}{
\begin{tabular}{rc c rr c rr c rr c rr}
  \hline
  	&					&& \multicolumn{5}{c}{UZIP}  && \multicolumn{5}{c}{MZIP$^{\dag}$}\\ \cline{4-8} \cline{10-14}
  	&					&& \multicolumn{2}{c}{Binary}	&& \multicolumn{2}{c}{Count} && \multicolumn{2}{c}{Binary} && \multicolumn{2}{c}{Count}\\ \cline{4-5} \cline{7-8} \cline{10-11} \cline{13-14}
	& 	True				&& $\alpha_{j,1}$ & 		 	&& $\beta_{j,1}$ & && $\alpha_{j,1} | \delta_{j,1}=1$ & $\delta_{j,1}$ 	&& $\beta_{j,1} | \gamma_{j,1}=1$ & $\gamma_{j,1}$ \\ 
 $j$	& $\alpha_{j,1}, \beta_{j,1}$ && Est (SE) 	& $m_{\alpha, j}$&& Est (SE) 					& $m_{\beta, j}$ 		&& PM (SD) 				& PM  			&& PM (SD) &					 PM\\
  \hline
1  & ~0.05 && 0.05 (0.04) & 0.02 &  & 0.042 ($0.002$) & 0.95 &  & 0.06 (0.04)   & 0.04 &  & 0.03 (0.02)   & 0.03 \\
  2 & ~0.10 && 0.11 (0.04) & 0.14 &  & 0.088 ($0.002$) & 1.00 &  & 0.11 (0.04)   & 0.43&  & 0.09 (0.02)   & 1.00  \\
  3 & ~0.15 && 0.15 (0.05) & 0.63 &  & 0.142 ($0.002$) & 1.00 &  & 0.15 (0.04)   & 0.95 &  & 0.15 (0.02)   & 1.00 \\
  4 & ~0.20 && 0.20 (0.05) & 0.91 &  & 0.191 ($0.002$) & 1.00 &  & 0.20 (0.04)   & 1.00 &  & 0.19 (0.02)   & 1.00 \\
  5 & ~0.25 && 0.25 (0.05) & 0.99 &  & 0.233 ($0.002$) & 1.00 &  & 0.24 (0.04)   & 1.00&  & 0.24 (0.02)   & 1.00  \\
  6 & -0.05 && -0.04 (0.04) & 0.01 &  & -0.046 ($0.002$) & 0.97 &  & -0.06 (0.04)   & 0.04 &  & -0.06 (0.02)   & 0.23 \\
  7 & -0.10 && -0.11 (0.04) & 0.25 &  & -0.104 ($0.002$) & 0.99&  & -0.11 (0.04)   & 0.36&  & -0.11 (0.02)   & 1.00  \\
  8 & -0.15 && -0.14 (0.05) & 0.55 &  & -0.143 ($0.002$) & 1.00&  & -0.15 (0.04)  & 0.95 &  & -0.15 (0.02)   & 1.00 \\
  9 & -0.20 && -0.20 (0.05) & 0.95 &  & -0.194 ($0.002$) & 1.00&  & -0.20 (0.04)   & 1.00 &  & -0.20 (0.02)   & 1.00 \\
  10 & -0.25 && -0.25 (0.05) & 0.99 &  & -0.251 ($0.002$) & 1.00&  & -0.24 (0.04)   & 1.00&  & -0.25 (0.02)  & 1.00  \\
  11 & ~0.00 && -0.01 (0.04) & 0.00 &  & 0.001 ($0.002$) & 0.82&  & -0.01 (0.04)   & 0.01&  & 0.00 (0.01)   & 0.00  \\
  12 & ~0.00 && 0.02 (0.04) & 0.00 &  & 0.001 ($0.002$) & 0.84&  & 0.02 (0.04)   & 0.02 &  & 0.00 (0.01)   & 0.00 \\
  13 & ~0.00 && -0.01 (0.04) & 0.00 &  & -0.004 ($0.002$) & 0.89&  & -0.00 (0.04)   & 0.01&  & -0.00 (0.01)   & 0.00  \\
  14 & ~0.00 && -0.00 (0.04) & 0.00 &  & -0.002 ($0.002$) & 0.82&  & -0.00 (0.04)   & 0.01 &  & -0.00 (0.01)   & 0.00 \\
  15 & ~0.00 && 0.00 (0.04) & 0.00 &  & -0.000 ($0.002$) & 0.93&  & -0.00 (0.04)   & 0.01&  & -0.00 (0.01)   & 0.01  \\
  16 & ~0.00 && -0.00 (0.04) & 0.00 &  & -0.004 ($0.002$) & 0.80&  & -0.00 (0.04)   & 0.01 &  & -0.00 (0.01)   & 0.00 \\
  17 & ~0.00 && 0.00 (0.04) & 0.00 &  & -0.005 ($0.002$) & 0.85&  & 0.00 (0.04)   & 0.01 &  & -0.00 (0.01)   & 0.00  \\
  18 & ~0.00 && 0.01 (0.04) & 0.00 &  & 0.007 ($0.002$) & 0.96&  & 0.00 (0.04)   & 0.01&  & -0.00 (0.01)   & 0.00  \\
  19 & ~0.00&& 0.00 (0.04) & 0.01 &  & -0.003 ($0.002$) & 0.85 &  & -0.00 (0.04)   & 0.01 &  & 0.00 (0.01)   & 0.00 \\
  20 & ~0.00 && -0.01 (0.04) & 0.00 &  & 0.002 ($0.002$) & 0.83&  & -0.00 (0.04)   & 0.01&  & -0.00 (0.01)   & 0.00  \\
   \hline
   \multicolumn{14}{l}{\footnotesize$\dag$ The medians of the maximum likelihood estimate (Est) and standard error (SE) of $\alpha_{j,1}$ and $\beta_{j,1}$, the proportion that}\\   
   \multicolumn{14}{l}{\footnotesize ~  the covariate is selected for the j$^{\textrm{th}}$ outcome ($m_{\alpha, j}$, $m_{\beta, j}$) are calculated.}\\
   \multicolumn{14}{l}{\footnotesize$*$ The empirical standard deviations of Est($\beta_{j,1}$) range between 0.036 and 0.054. (not presented in the table)}\\      
   \multicolumn{14}{l}{\footnotesize$\ddag$ The medians of the posterior means (PM) and posterior standard deviation (SD) of $\alpha_{j,1}$ and $\beta_{j,1}$ (conditioning on}\\      
   \multicolumn{14}{l}{\footnotesize ~ $\delta_{j,1}=1$ and $\gamma_{j,1}=1$, respectively), the medians of the posterior means of $\delta_{j,1}$ and $\gamma_{j,1}$ (marginal posterior probability}\\   
   \multicolumn{14}{l}{\footnotesize ~ of inclusion) are computed. The hyperparameters $\omega_{\alpha}=\omega_{\beta}$ are set to 0.1.}\\   
   \multicolumn{14}{l}{\footnotesize NOTE: Throughout values are based on results from 100 simulated datasets.}\\      
\end{tabular}}
\end{table}

\subsection{Results}

We focus the presentation of results in this section on the MZIP model with $\omega_{\beta, 1}$=$\omega_{\alpha, 1}$=$0.1$. This is to demonstrate the improvement gained by the proposed multivariate approach over an analogous univariate method, while implementing the fairest comparison to the univariate method. When the values of the overall prior inclusion probabilities ($\omega_{\beta, 1}$ and $\omega_{\alpha, 1}$) increased from 0.1 to 0.5, the MZIP model tended to select one to two more variables on average, yielding higher TPR and NPV but also a bit higher FPR and lower PPV in both parts of the model (Table \ref{sim:tbl:roc}). When the outcome variables were uncorrelated (Scenario VI), the variable selection capability for the MZIP model with $\omega_{\beta, 1}$=$\omega_{\alpha, 1}$=$0.1$ was almost the same as that of UZIP model. 

Across scenarios in which the outcomes were correlated (I-III), the binary part of the MZIP model was more sensitive than in the UZIP approach (Table \ref{sim:tbl:roc}), for example, with TPR=61\% versus 54\% in Scenario II, respectively. The TPR in UZIP models was insensitive to the strength of correlation among outcomes, whereas the MZIP TPR increased to 67\% in Scenario I, in which there was a stronger correlation among outcomes associated with the covariate. Both the UZIP and the MZIP methods successfully identified the covariate associations for the four outcomes with the largest effect sizes ($|\alpha_{j,1}|\geq 0.20$; Table \ref{sim:tbl:est}). However, the MZIP model performed much better at detecting smaller-magnitude associations: when $|\alpha_{j,1}| = 0.10$ (outcomes 2 and 7), associations were correctly included in 40\% and 20\% of MZIP and UZIP models, respectively; the corresponding inclusion rates were 95\% and 60\% when $|\alpha_{j,1}| = 0.15$ (outcomes 3 and 8). 

The multivariate approach yielded much more substantial improvement for the count part of model when outcomes were correlated (Table \ref{sim:tbl:roc}). Even controlling the FDR, the univariate approach generally exhibited inflated type I error, as high as 86\% across Scenarios I-V. In contrast, across all scenarios the MZIP model had a low probability of false discovery (FPR$<1$\%) while also exhibiting high TPR that ranged from 73\% to 83\% in Scenarios I-IV. The relatively poor performance of the UZIP method is not due to bias, since the estimated association for outcomes unassociated with the covariate ($\beta_{j,1}$, $j$=11,\ldots,20) were very close to zero (Table \ref{sim:tbl:est}). However, the medians of the asymptotic standard errors (SE) for $\hat{\beta}_{j,1}$ were ~20 times smaller than the empirical standard deviations of $\hat{\beta}_{j,1}$ (rang between 0.036 and 0.054) (Table \ref{sim:tbl:est}), was also observed in Scenario II-V (Supplementary Material D). Thus, the univariate approach appears to perform poorly in estimation of the standard errors for count model parameters when outcomes are correlated. Consequently, the estimated confidence intervals are too narrow. 

In the \emph{null case} (Scenario V), both approaches successfully excluded the covariate for all outcomes for the binary part of the model even when outcomes were strongly correlated; the UZIP model exhibited a high false discovery rate (87\%) for the count part of the model. 

We ran extensive additional simulations to explore other factors (detailed in Table E.1 and E.2 in Supplementary Material E), including a larger number of outcomes ($q$=50), a lower signal density (4$\sim$5\%), a smaller sample size ($n$=150) and negative correlations. Briefly, the results were similar to those described above, with the MZIP performing much better for variable selection and the UZIP exhibiting inflated type I error. We also compared the proposed MZIP to a univariate non-parametric method, the Wilcoxon rank sum test. Although type I error was well controlled in the univariate non-parametric method, substantially higher power was achieved by the MZIP. 

To summarize, compared with the univariate approach, the proposed multivariate method improved upon the UZIP's performance for the binary part of the model by maintaining type I error while boosting the ability to identify true associations under the simulated settings. For the count part of the model, there were some scenarios in which the power of UZIP was higher than with the MZIP approach. This higher power of UZIP was at a cost of a highly inflated false discovery rate, whereas the MZIP FPR was $<1$\%. Performance of the MZIP model was enhanced by increasing the prior inclusion probabilities, $\omega_{\beta, k}=\omega_{\alpha, l}=0.5$. The TPR then exceeded 80\% in all non-null scenarios and FPR remained $<4$\% across all scenarios.

\section{Application to Pediatric HIV/AIDS Cohort Study Data}

The proposed MZIP method was originally motivated by research into whether caries-associated bacteria differ in PHIV and PHEU youth \citep{moscicki2016burden, ryder2017prevalence}. The 254 subjects were age 10 to 22 years at the time of an oral health examination done from September 2012 to January 2014. Subgingival dental plaque samples were collected at two preselected sites and excluded if participants had antibiotic exposure in the previous three months. DNA was isolated from plaque specimens and 16S rDNA sequenced \citep{caporaso2011global, gomes2015microbiomes}. The sequencing reads were trimmed, filtered, and grouped using the DADA2 pipeline, and reads matched to the curated Human Oral Microbiome Database (99.9\% of reads matched to the species or genus level) \citep{dewhirst2010human}. Each subject had a count (number of sequencing reads) for each taxon identified in the study.

\subsection{Analysis Details and Prior Specifications}

We focused our analysis on $q=44$ taxa: 14 known caries-associated species \citep{aas2008bacteria} and any additional species that were highly correlated with them in this dataset  (Figure \ref{sim:fig:pred} (a)). HIV infection status ($x_{i1}$=$z_{i1}$; 0, uninfected; 1, infected) was the covariate of primary interest, with adjustment for participants' age ($x_{i2}$=$z_{i2}$) without performing variable selection on it (i.e., it was ``forced" into the model). To account for sequencing depth variation across samples, the total number of sequencing reads was included as an offset in the count model. 

We fit the proposed MZIP model and the UZIP model to PHACS data. For the Bayesian MZIP approach, we set the hyperparameters, ($\bfmu_{\beta_0}$, $\bfmu_{\alpha_0}$, $a_{\beta_0}$, $b_{\beta_0}$, $a_{\alpha_0}$, $b_{\alpha_0}$, $v_{\beta, j}$, $v_{\alpha, j}$, $a_{\beta, k}$, $b_{\beta, k}$, $a_{\alpha, l}$, $b_{\alpha, l}$, $\Psi_0$, $\rho_0$), to the same values as in Section \ref{sec:simulation}. Since performance of the MZIP was improved when the prior inclusion probabilities were increased from 0.1 to 0.5 in the simulation studies, we set $\omega_{\beta, k}=\omega_{\alpha, l}$=0.5. We ran two independent MCMC chains for 2 million iterations, each with the first half taken as burn-in. We assessed convergence of the MCMC sampler by plotting traces of the MCMC scans for each parameter. A visual assessment of convergence to the stationary distribution was carried out by overlaying plots for the two MCMC chains. 

We also calculated the posterior median with 95\% credible intervals for the marginal IDR for MZIP (described in Section \ref{sec:IDR}). We age-adjusted the estimate by using the mean value, 16 years. 

\subsection{Results}
 
\begin{table}[ht]
	\caption{Estimated regression parameters and inclusion probabilities for the five species identified as associated with HIV infection by using a multivariate zero-inflated Poisson (MZIP)$^{\dag}$ and the univariate zero-inflated Poisson (UZIP)$^{\ddag}$ models.\label{tbl:plq:rp}}
\centering
\scalebox{0.7}{
\begin{tabular}{c c cc r cc r c}
  \hline
						&	& \multicolumn{2}{c}{Binary}					&&  \multicolumn{2}{c}{Count}		&& \multirow{2}{*}{IDR$^{*}$} \\   \cline{3-4}\cline{6-7}
						&	& $\alpha_{j,1}|\delta_{j,1}=1$ 	& \multirow{2}{*}{IP} && $\beta_{j,1}|\gamma_{j,1}=1$ &\multirow{2}{*}{IP} 	& & 	 \\
						&	& PM (95\% CI) 			&   				&& PM (95\% CI) 			&  	&& PM (95\% CI)\\ 
  \hline
	&\emph{Actinomyces graevenitzi} 						& -0.39 (-0.67, -0.09) & \bf{0.54} 	&& 0.25 (-0.22, 0.72)  & 0.11 	&& 0.96 (0.57, 1.08) \\ \\
	&  \emph{Fusobacterium periodonticum} 					& 0.16 (-0.60, 0.48) & 0.07 	&& -0.43 (-0.66, -0.14) & \bf{0.93} 	&& 0.66 (0.53, 0.98) \\ \\
MZIP & \emph{Lachnoanaerobaculum orale} 					& 0.10 (-0.17, 0.33) & 0.03 	&& -0.67 (-1.15, -0.25) & \bf{0.95} 	&& 0.52 (0.24, 0.86) \\ \\
 	& \emph{Leptotrichia} sp oral taxon 215 				& -0.44 (-0.75, -0.17) & \bf{0.63}  && -0.04 (-0.19, 0.20) & 0.02 	&& 0.86 (0.74, 1.00) \\ \\
 	& \emph{Veillonella} genus NOI$^{\S}$ 						& -0.04 (-0.60, 0.45) & 0.05 	&& -0.37 (-0.60, -0.12) & \bf{0.88} 	&& 0.71 (0.57, 1.00) \\ \\
   \hline 
   \hline
   						&	& $\alpha_{j,l} $ 	& && $\beta_{j,k}$ & 	& & 	 \\
						&	& Est (95\% CI$_F$) 			&   				&& Est (95\% CI$_F$) 			&  	&& \\ 
  \hline
	&\emph{Actinomyces graevenitzi}						& -0.27 (-0.89, 0.35) &&& -0.46 (-0.69, -0.23) &&& \\ \\
 	& \emph{Fusobacterium periodonticum} 					& -0.08 (-0.93, 0.77) &&& -0.80 (-0.84, -0.75) &&&\\ \\
UZIP & \emph{Lachnoanaerobaculum orale} 					& 0.31 (-0.38, 0.99) &&& -2.58 (-2.79, -2.36) &&&\\ \\
 	& \emph{Leptotrichia} sp oral taxon 215 				& -0.41 (-1.04, 0.21) &&& -0.26 (-0.41, -0.11) &&&\\ \\
 	& \emph{Veillonella} genus NOI  						& 0.20 (-1.02, 1.41) &&& -0.24 (-0.27, -0.21) &&&\\  \\
 \hline
  \multicolumn{9}{l}{\footnotesize$^{\dag}$ In the MZIP model, the posterior median (PM) and 95\% credible interval (CI) of regression parameters and incidence density ratio (IDR) are}\\
  \multicolumn{9}{l}{\footnotesize ~ computed.}\\  
  \multicolumn{9}{l}{\footnotesize$^{\ddag}$ In the UZIP model, the maximum likelihood estimates (Est) and 95\% confidence intervals (CI$_F$) of regression parameters are computed.}\\
 \multicolumn{9}{l}{\footnotesize$^{*}$ Adjusted for individuals with age of 16 years.}\\
  \multicolumn{9}{l}{\footnotesize$^{\S}$ not otherwise identified}\\
\end{tabular}}
\end{table}


\begin{figure}[hp]
	\centering
	\includegraphics[width=5in]{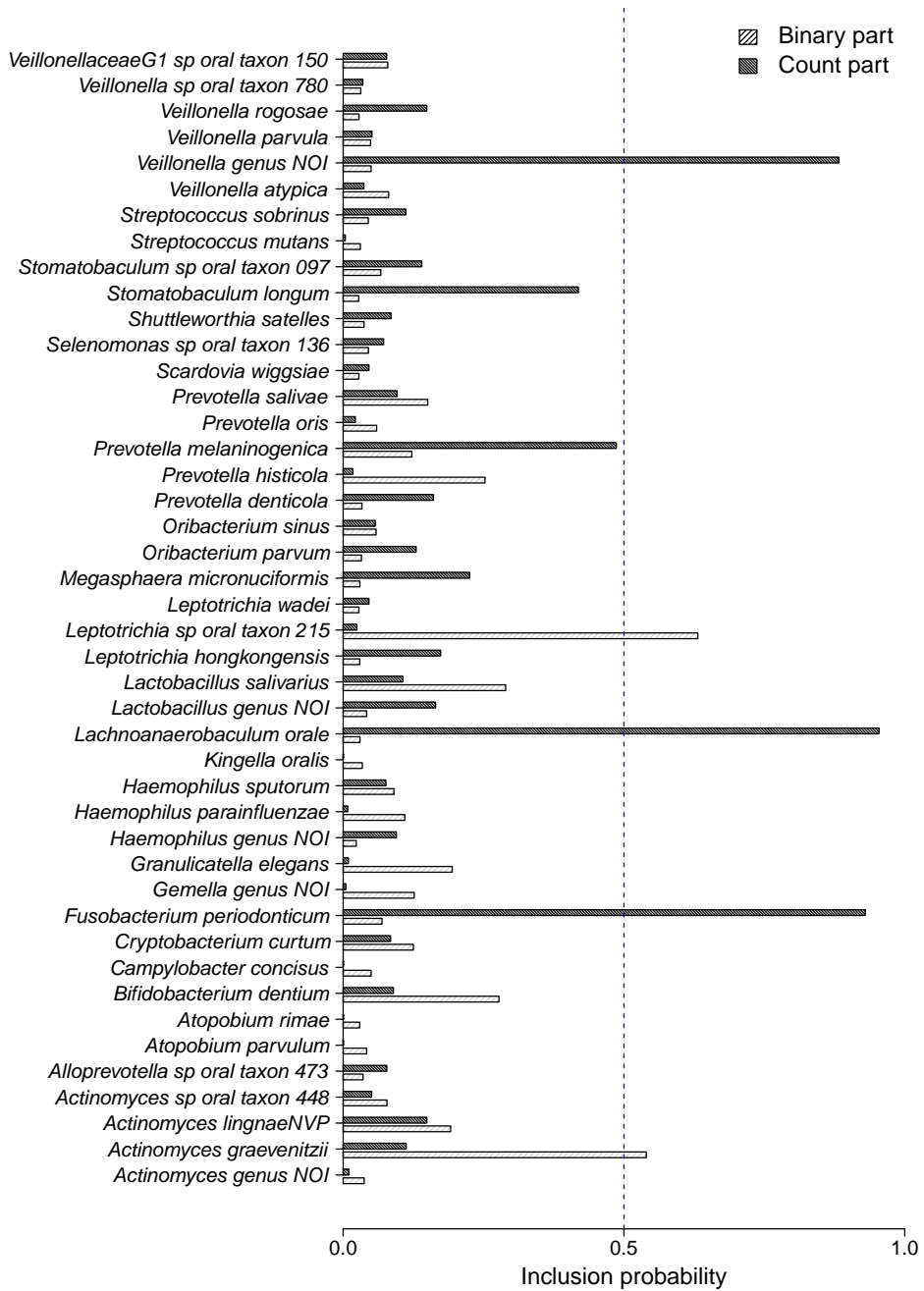}
	\caption{Analysis of Pediatric HIV/AIDS Cohort Study (PHACS) data: marginal posterior inclusion probabilities for the HIV status covariate in relation to excess zeros and counts of 44 microbial species as estimated via the proposed multivariate zero-inflated Poisson (MZIP) model. We adjust for participantsÕ age, a potential confounder, but do not perform variable selection on it. ``NOI" stands for ``not otherwise identified".  \label{fig:thw:gamma}}
\end{figure}

\begin{figure}[hp]
\centering
\includegraphics[width=6.5in]{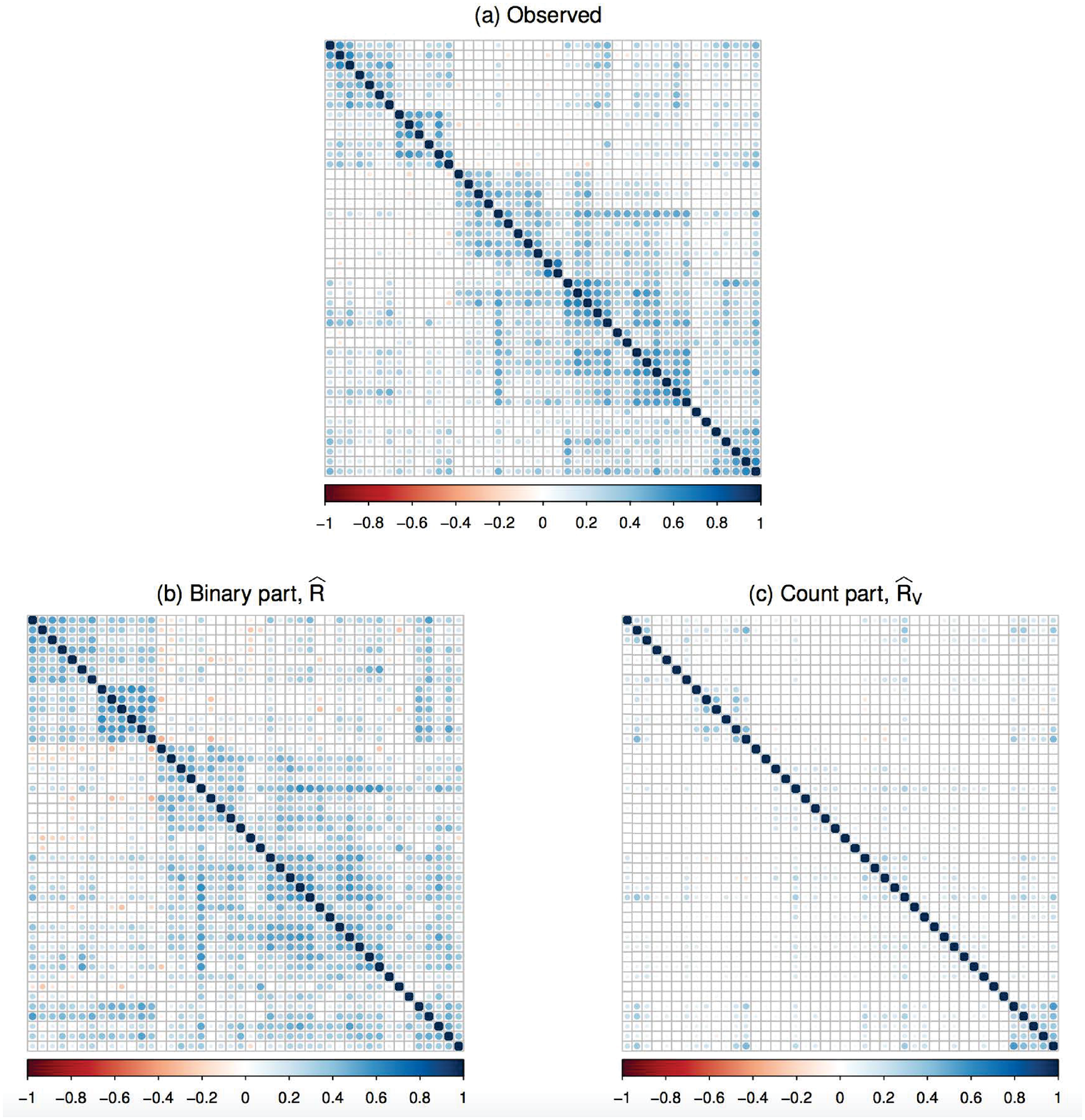} \\
\caption{Observed and estimated correlations among counts of 44 microbial species in the Pediatrics HIV/AIDS Cohort Study (PHACS): (a) Empirical correlations calculated by cor($\log(y + 1)$); (b) Posterior median of $R$; (c) Posterior median of $R_V$, calculated based on posterior samples of $\Sigma_V$ from the proposed multivariate zero-inflated Poisson (MZIP) model fit.}
\label{sim:fig:pred}
\end{figure}

With a marginal posterior probability cutoff of 0.5, the MZIP method identified three species (\emph{Fusobacterium periodonticum}, \emph{Lachnoanaerobaculum orale}, and \emph{Veillonella} genus NOI) for which counts were associated with HIV infection among ``susceptible" individuals, and it selected two species (\emph{Actinomyces graevenitzii} and \emph{Leptotrichia} sp oral taxon 215) for which the probability of participants being susceptible to these two species was associated with HIV infection (Figure \ref{fig:thw:gamma}). In contrast, the frequentist UZIP analysis with 95\% confidence intervals identified 39 species whose levels were associated with HIV infection in the susceptible population, including the three associations identified by MZIP. The UZIP analysis identified no species for which the probability of being ``susceptible" to that species was associated with HIV infection. Based on the methods' relative performance in the simulation study, it is plausible that the UZIP's lack of accounting for outcomes' correlation patterns, which are complex (Figure \ref{sim:fig:pred} (a)), grossly inflated the type I error rate in the count model and may also have decreased sensitivity of the zero model. 

Indeed, comparing the estimated associations between HIV infection and the five taxa selected based on the MZIP model (Table \ref{tbl:plq:rp}), the uncertainty associated with the count part of UZIP model was much smaller than that generated by the MZIP method. As with the simulation, this may have accounted for the detection of 39 associations, many of which are presumably false positive associations. Because of how we calculated the IDR estimate, it has an age-specific interpretation. For example, 16 year-old youth, PHIV youth had 34\% (95\% credible interval 2\%, 47\%) and 48\% (95\% credible interval 14\%, 76\%) lower abundance of \emph{Fusobacterium periodonticum} and \emph{Lachnoanaerobaculum orale} species, respectively, compared with PHEU youth.

The proposed MZIP model captures within-subject dependence among multiple outcomes via two correlation components, $R$ and $\Sigma_V$. The dependence patterns arising from the empirical correlations appear to reflect, strongly, the correlation structure predicted by the binary component of the proposed MZIP (Figure \ref{sim:fig:pred}). This implies that in these data, the presence of taxa is more structured than their counts. This result is not attributable to smoothing of the empirical correlation from having added 1 to every count, because the results were not sensitive to changes of this value to 0.5, 0.1, and 0.01. The model also provides an opportunity to quantify and compare the contribution of zero inflation versus other sources of overdispersion to microbiome taxon abundance (see Supplementary Material F for posterior estimates of $\Sigma_{Vj,j}$ and $\alpha_{0,j}$).

\section{Discussion}

We have described the development of a new Bayesian variable selection method that addresses challenges arising in the analysis of microbiome sequencing data: excess zero counts and high-dimensional outcomes with a complex association structure. Applying the proposed multivariate approach led to the identification of two species for which the probability of being susceptible to those species was associated with HIV infection; these associations did not meet FDR thresholds when the existing univariate approach was applied. In addition, based on the estimated induced marginalized IDR under the proposed model, another two species were less abundant in HIV-infected youth aged 16 years compared with PHEU youth of the same age. 

One might question how realistic are these analyses when they are adjusted for only one confounder, age. Some reassurance might be provided from the observation that in univariate analyses that included additional confounders (e.g., sex, race, and dental visit in previous year as a marker of oral hygiene), inference was not greatly altered compared with models including only HIV status and age. We are continuing to study performance in a range of datasets, including more complete confounder adjustment. We are also working to scale up the proposed method in the number of endpoints, as discussed further below, and also in the number of covariates, both of which are required for integrated omics analyses. 

The simulation study demonstrated superior performance of the proposed MZIP approach over the existing UZIP method when outcomes were correlated. The sample size was small enough that asymptotic assumptions under the frequentist-based UZIP model did not hold. This affected estimation of the asymptotic variance of the regression parameters for the count model, which was not a limitation for the proposed multivariate Bayesian approach. This difference in performance is primarily because i) for small data settings, estimation is generally more stable with Bayesian approaches, which exploit information from both the observed data likelihood and prior distribution; and ii) the MZIP method uses information not only on the mean model but also from the structure of covariance among outcomes.

We used a multivariate probit model for the binary part of the MZIP mixture model. An alternative is to assume a multivariate logistic distribution for $\bfw_i$ \citep{o2004bayesian}, for which posterior computation can be facilitated based on a data augmentation algorithm \citep{albert1993bayesian}. However, initial numerical studies using the latter approach resulted in prohibitively slow mixing of the MCMC algorithms due to sparseness of data, even for data with $q$=10 and assuming an unstructured covariance pattern. This is because the multivariate logistic model specification requires the estimation of $n$ more latent parameters than does the multivariate probit model. Thus, the multivariate probit model in the MZIP proved to be much more computationally tractable. 

We have presented the model in its most general form that allows the importance of each covariate, as well as the correlation structure among the multivariate outcomes, to vary across the binary and the count components of the model. This gives the user maximal flexibility and provides evidence on how  a covariate is associated with each response, i.e. with more zeros or higher counts. The question arises whether the complexity of the model is necessary or whether simpler models should suffice. Analysis of the motivating data suggests that different correlation structures were needed in this case. It is difficult to provide a general answer to this question until we have had more opportunity to apply it a range of datasets and compare results with those obtained in simpler models. One would not be able to make this comparison if the most general model is not available as a basis for comparison. Yet, simpler models might well be useful in other datasets. 

Thus, the software implementation of the proposed approach offers more parsimonious versions of the model, simplified by imposing additional restrictions regarding the model parameters. For example, the two model parts can be forced to have one common variable selection indicator by setting $\gamma_{j,k}=\delta_{j,l}$. In practice, such a restriction might facilitate implementation by providing a single vector of variable selection indicators, i.e. one list as to which species are associated with each covariate. A different assumption is that both model parts share the same covariance pattern ($R=R_V$), which will greatly reduce the number of parameters to be estimated and thus the computational complexity in the MCMC algorithm, especially for data with large $q$. In our initial analysis, fitting this restricted model to the PHACS data yielded unreliable estimates of $R$, because the assumption that $R=R_V$ is violated in these data (Figure \ref{sim:fig:pred}). Again, the restricted model may serve well in other datasets. Therefore, we made available the algorithms to implement both types of simpler MZIP models as options in the \texttt{mBvs} package.

There are several ways the proposed framework could be extended. First, marginalized zero-inflated models have recently been developed so that inference can be made on the marginal mean of the sampled population via a set of unified regression coefficients \citep{leann2015marginalized,tabb2016marginalized}. The unified regression coefficients have better interpretability, as the marginalized models do not require the additional steps described in Section \ref{sec:IDR} to address the dependence of parameter values on $\bfx_{i,(-k)}$. In some applications, it may be more appropriate to interpret the two sets of regression coefficients separately, yet there also may be other applications for which interpretability of regression parameters would be enhanced by adopting a marginalized model within the proposed multivariate Bayesian variable selection method. Marginalization may also provide more stable model fitting. Second, although we focused data analysis on a preselected subset of species in the application, often microbiologists' goal is to perform whole-community oral microbiome analysis, which generally involves several hundred taxa. We are currently working to address the computational issues arising from an even higher-dimensional parameter space. Because the complexity mainly results from the flexible unstructured covariance model, we propose to scale up the proposed MZIP method by adopting alternative correlation structures that can flexibly accommodate potentially complicated patterns among hundreds of taxa. A final possibility is to study the interaction between the marginal posterior probability cutoff and the prior inclusion probability in controlling the FDR at the desired level under the proposed model.

In conclusion, the proposed framework gives researchers valid and powerful statistical tools to overcome major methodological barriers in microbiome sequencing data analysis. Beyond the study of the human microbiome, the methods, software, and guidance from simulation studies in this work will be useful in any field requiring analysis of multivariate zero-inflated count data.

\bigskip
\begin{center}
{\large\bf SUPPLEMENTARY MATERIALS}
\end{center}

\begin{description}

\item[R-package `\texttt{mBvs}':] R-package \texttt{mBvs} contains codes to implement proposed Bayesian framework described in the article. The package is currently available in CRAN (https://cran.r-project.org/web/packages/mBvs).

\end{description}

\bibliographystyle{apalike}
\bibliography{MZIP}

\end{document}